\documentclass[twocolumn,showpacs,pra,superscriptaddress,nofootinbib]{revtex4-1}
\usepackage{graphicx}
\usepackage{amsmath}

\newcommand{\bra}[1]{\mbox{$\left\langle #1 \right|$}}
\newcommand{\ket}[1]{\mbox{$\left| #1 \right\rangle$}}

\begin{document}

\title{Practical Quantum Digital Signature}
\author{Hua-Lei Yin}
\email{hlyin@mail.ustc.edu.cn}
\author{Yao Fu}
\email{yaofu@mail.ustc.edu.cn}
\author{Zeng-Bing Chen}
\email{zbchen@ustc.edu.cn}
\affiliation{Hefei National Laboratory for Physical Sciences at Microscale and Department
of Modern Physics, University of Science and Technology of China, Hefei,
Anhui 230026, China}
\affiliation{The CAS Center for Excellence in QIQP and the Synergetic Innovation Center
for QIQP, University of Science and Technology of China, Hefei, Anhui
230026, China}
\date{\today}

\begin{abstract}
Guaranteeing non-repudiation, unforgeability as well as transferability of a signature is one of the most vital safeguards in today's e-commerce era. Based on fundamental laws of quantum physics, quantum digital signature (QDS) aims to provide information-theoretic security for this cryptographic task.
However, up to date, the previously proposed QDS protocols are impractical due to various challenging problems and most importantly, the requirement of authenticated (secure) quantum channels between participants.
Here, we present the first quantum digital signature protocol that removes the assumption of authenticated quantum channels while remaining secure against the collective attacks.
Besides, our QDS protocol can be practically implemented over more than 100 km under current mature technology as used in quantum key distribution.
\end{abstract}

\pacs{03.67.Dd, 03.67.Hk, 03.67.Ac}

\maketitle

\section{Introduction}

Digital signatures aim to certify the provenance and identity of a message as well as the authenticity of a signature. They are widely applied in e-mails, financial transactions, electronic contracts, software distribution and so on.
However, relying on mathematical complexities, classical digital signature schemes  become vulnerable to quantum computing attacks.
Though there are some classical unconditionally secure signature schemes \cite{chaum:1991:unconditionally,shikata:2002:security,swanson:2011:unconditionally,Wallden:2015:QDS,Arrazola:2015:Multi}, a resource-expensive assumption exists therein, namely, the secure classical channels between the participants.
Quantum digital signature (QDS) bases on fundamental laws of quantum physics to guarantee its information-theoretic security.
Since Gottesman and Chuang proposed the first QDS protocol \cite{Gottesman:2001}, a quantum version of Lamport's scheme \cite{Lamport:1979}, the following problems appear therein: ($P1$) requiring the authenticated quantum channels, ($P2$) the preparation and transmission of complex one-way function quantum states, ($P3$) requiring long-term quantum memory, and ($P4$) other challenging operations, such as performing SWAP test on the states. A novel approach uses linear optics and photon detectors to circumvent the requirement for quantum memory and complex state preparation \cite{Dunjko:2014:Quantum} and replaces the SWAP test with an optical multiport \cite{clarke:2012:experimental}, yet leading to another challenging technology---long-distance stabilization of the Mach-Zehnder interferometer \cite{Collins:2014:Realization}.
Up to date, an obviously serious problem left for a feasible QDS protocol is the impractical assumption of the authenticated quantum channels between participants being available.

The authenticated quantum channels are equivalent to the secure quantum channels that do not allow any eavesdropping. Recall that while quantum key distribution (QKD) \cite{RevModPhys:09:Scarani,lo:2014:secure} requires authenticated classical channels, it does not require authenticated (secure) quantum channels because of potential eavesdropping. Actually, guaranteeing secure key distribution without authenticated quantum channels is exactly the goal and definition of QKD. Therefore, the assumption of authenticated quantum channels has to be eliminated in a QDS to be of practical value.
In this paper, using single-photon qubit state and phase-randomized weak coherent states, we propose the first QDS protocol that eliminates the impractical assumption of secure quantum channels.
Therefore, in respect to tackling the security problem merely requiring the authentication of classical communication, the basic assumptions underlying our QDS protocol are similar to that of QKD \cite{RevModPhys:09:Scarani,lo:2014:secure} and multiparty quantum communication (quantum secret sharing) \cite{Hillery:1999:QSS,YF:2015:MQC}.

We say a digital signature protocol is secure if it satisfies \cite{swanson:2011:unconditionally}: unforgeability, non-repudiation and transferability. Unforgeability means that a given piece of message indeed comes from the signer and remains intact during transmission, namely, no one can forge a valid signature that can be accepted by other honest recipients.
Non-repudiation means that once the signer signs a message, he/she cannot deny having signed it.
Transferability means that when an honest recipient of the signed message accepts a signature, other honest recipients will accept it as well.
We define a QDS scheme to have $\varepsilon_{\textrm{unf}}$-unforgeability, representing that the probability for an adversary to create a valid signature is not greater than $\varepsilon_{\textrm{unf}}$. Similarly, we say a protocol is with $\varepsilon_{\textrm{nor}}$-non-repudiation when the probability for the signer to repudiate a legitimate signature is not greater than $\varepsilon_{\textrm{nor}}$. Thereby, a QDS protocol is defined to have $\varepsilon_{\textrm{sec}}$-security when it satisfies both $\varepsilon_{\textrm{unf}}$-unforgeability and $\varepsilon_{\textrm{nor}}$-non-repudiation, with $\varepsilon_{\textrm{unf}}+\varepsilon_{\textrm{nor}}\leq\varepsilon_{\textrm{sec}}$.
Considering robustness, $\varepsilon_{\textrm{rob}}$ is the probability that the protocol will be aborted even with the absence of an adversary.

In this paper, we consider a simple and most important case with three participants, i.e., one signer and two recipients. Then the property of transferability becomes equivalent to non-repudiation \cite{Gottesman:2001,Dunjko:2014:Quantum}. The QDS protocols using two copies of single-photon states and decoy-state method are proposed in Sec. II and Sec. IV, with the security analyzed in Sec. III.

\section{QDS with two copies of single-photon states}

There are three stages when implementing our three-participant QDS protocol, namely, the distribution stage, the estimation stage and the messaging stage.
We introduce a two-photon six-state QDS protocol to illustrate our basic idea.
There are six single-photon quantum states, $\ket{H}$, $\ket{V}$, $\ket{\pm}=(\ket{H}\pm\ket{V})/\sqrt{2}$, $\ket{R}=(\ket{H}+i\ket{V})/\sqrt{2}$ and $\ket{L}=(\ket{H}-i\ket{V})/\sqrt{2}$. The six states can be arranged into twelve sets $\{\ket{H},\ket{+}\}$, $\{\ket{+},\ket{V}\}$, $\{\ket{V},\ket{-}\}$, $\{\ket{-},\ket{H}\}$, $\{\ket{H},\ket{R}\}$, $\{\ket{R},\ket{V}\}$, $\{\ket{V},\ket{L}\}$, $\{\ket{L},\ket{H}\}$, $\{\ket{+},\ket{R}\}$, $\{\ket{R},\ket{-}\}$, $\{\ket{-},\ket{L}\}$, $\{\ket{L},\ket{+}\}$, where the first state of each set represents logic 0 and the second logic 1.

\emph{The distribution stage}: For each possible future message $m=0$ and $m=1$, Alice prepares two copies of a sequence of $N$ single-photon quantum states.
For each quantum state, Alice randomly chooses one of the twelve sets and generates one of two non-orthogonal states in the set. Afterwards, she sends one copy to Bob and the other to Charlie through insecure (unauthenticated) quantum channels. For each quantum state, Bob and Charlie randomly and independently perform a polarization measurement with one of the three bases $\{Z, X, Y\}$, and store the corresponding classical bit.
Bob and Charlie will announce the result if their detectors have no click, and then Alice, Bob and Charlie will discard all the corresponding data and keep the left $M$ bits. For each quantum state, Alice announces from which set she selects the state through the authenticated classical channels. Bob (Charlie) compares his measurement outcomes with the two states. If
his measurement outcome is orthogonal to one of the states, he concludes that the other state has been sent, which represents a conclusive result. Otherwise, he concludes that it is an inconclusive outcome. Let $P_{B}^{c}$ ($P_{C}^{c}$) be the probability that Bob (Charlie) has a conclusive result for each received quantum state; in the ideal case,  $P_{B}^{c}=P_{C}^{c}=P^{c}=1/6$. Note that Bob and Charlie do not announce whether they have a conclusive outcome.

\emph{The estimation stage}: The signer Alice chooses the desired recipient, for example Bob, who will be the authenticator in the messaging stage. Then Alice informs the other recipient, Charlie, to randomly choose $M_{t}$ bits as the test bits used to estimate correlation (if Charlie is the authenticator chosen by Alice, Alice will inform Bob to randomly choose test bits as well).
Charlie announce the location of test bits and Alice publicly announces the bit information of those test bits.  Bob (Charlie) calculates the mismatching rate
$e_{B}^{c}$ ($e_{C}^{c}$) of conclusive results from the test bits. When $e_{B}^{c}$ or $e_{C}^{c}$ gets too high, they announce to abort the protocol.
Besides, when $P_{B}^{c}$ or $P_{C}^{c}$ shows a big deviation from the ideal value $P^{c}=1/6$, they also announce to abort the protocol. Otherwise, Bob and Charlie announce the mismatching rate and the probability, $\{e_{B}^{c}, P_{B}^{c}\}$ and $\{e_{C}^{c}, P_{C}^{c}\}$, respectively. Alice, Bob and Charlie only keep $M_{u}$ untested bits, denoted by $S_{A}$, $S_{B}$ and $S_{C}$.

\emph{The messaging stage}:
To sign one-bit message $m$, Alice sends the message $m$ and the corresponding bit string $S_{A}$ to the authenticator, Bob. Bob checks the mismatching rate $P_{B}^{c}E_{B}^{c}$ between $S_{A}$ and $S_{B}$, where $E_{B}^{c}$ is the mismatching rate of the conclusive results. The inconclusive outcomes are considered to match Alice's announcement bits automatically.
If the mismatching rate $E_{B}^{c}\leq T_{a}$ ($T_{a}$ is the authentication security threshold), Bob accepts the message. Otherwise, he rejects it and announces to abort the protocol. After Bob accepts the message, he forwards it and the corresponding bit string $S_{A}$ to the verifier Charlie.
Charlie checks the mismatching rate $P_{C}^{c}E_{C}^{c}$ between $S_{A}$ and $S_{C}$, where $E_{C}^{c}$ is the mismatching rate of the conclusive results. If the mismatching rate $E_{C}^{c}\leq T_{v}$ ($T_{v}$ is the verification security threshold), Charlie accepts the forwarded message, otherwise he rejects it.

Note that the distribution stage is a quantum process, while the estimation and messaging stage are classical communication processes. The time interval between the distribution stage and estimation stage is arbitrary. The estimation stage is employed to estimate the parameters $T_{a}$ and $T_{v}$, which are used for the messaging stage. After the distribution stage, once Alice wants to sign the message, she will start the estimation stage and the messaging stage. As Alice is the signer, she can identify the one who is the desired recipient (namely the authenticator) before the estimation stage. Therefore, the roles of Bob and Charlie are equivalent, either of them can be the receiver of the message and forwards it to the other.

\section{security analysis}

In the three-participant QDS protocol, at most one participant can be an adversary, because the majority vote is usually used to resolve the dispute \cite{Dunjko:2014:Quantum,Arrazola:2015:Multi}.
Then, the only potential attack strategy can be either the repudiation of the signer or the forgery of the authenticator.

\begin{figure*}[tbh]
\centering \resizebox{13cm}{!}{\includegraphics{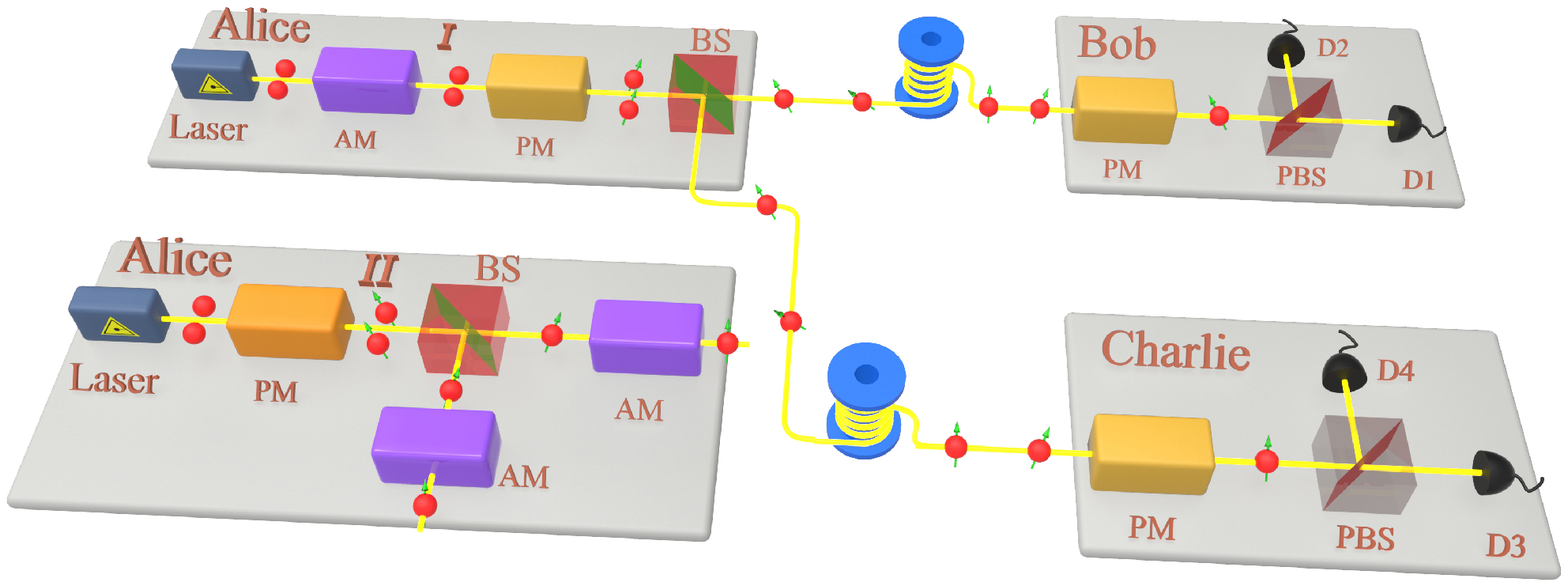}} \caption{(color online). Practical implementation of the three-participant QDS protocols.
Alice's setups for Scheme I (II) are shown respectively on the upper (lower) panel on the left. For the parts of Bob and Charlie, the setups are located on the right, which are shared by both schemes.
AM: amplitude modulator used to prepare decoy-state, PM: polarization modulator used to prepare polarization states or active basis selection, BS: 50:50 beam splitter used to prepare two copies of quantum states, PBS: polarization beam splitter, D1-D4: single-photon detectors. }
\label{f1}
\end{figure*}

We are now interested in why our scheme can prevent adversary's attack without authenticated quantum channels.
For Alice's repudiation attack, authenticated quantum channels cannot help her because the quantum states are prepared by herself. During the estimation stage, exploiting the random sampling theorem, one can estimate the correlation strength between Bob's and Charlie's received quantum states. Therefore, the SWAP test and symmetry operation of Bob's and Charlie's quantum states or classical bits can be removed in our protocol, which are realized by authenticated quantum channels \cite{Wallden:2015:QDS,Gottesman:2001,Dunjko:2014:Quantum,clarke:2012:experimental,Collins:2014:Realization,amiri:2015:unconditionally,Donaldson:2016:QDS} or secure classical channels \cite{Wallden:2015:QDS,Arrazola:2015:Multi,amiri:2015:secure} in other protocols.
For Bob's forgery attack, exploiting the random sampling theorem, Charlie can estimate the correlation strength between the quantum states Alice sends and those Charlie receives without authenticated quantum channels, whereas in previous protocols, authenticated quantum channels are required to guarantee that one copy of the quantum state Charlie receives has not been tampered with. Detailed security analysis are shown in Appendix A.

When Bob forges, Alice and Charlie are automatically regarded as honest.
A successful forgery means that the dishonest Bob forges (tampers) the message bit and Charlie accepts it.
Bob is unable to discriminate two copies of six-state (or four-state) without any error, since an unambiguous discrimination among $C$ linearly dependent qubit states is only possible when at least $C-1$ copies of the states are available \cite{Chefles:2001:Unambiguous}. For this reason, we can restrict Bob's forgery attack even when quantum channels are insecure.
We consider the case that Bob is restricted to collective forgery attack where his optimal strategy is to acquire the information of each quantum state Charlie receives as much as possible (exploiting quantum de Finetti theorem \cite{renner:2005:security}, we expect that our scheme could guarantee the security against coherent forgery attack, which should be studied in the future).
Thus, Bob's attack strategy in the two-photon six-state QDS protocol can be reduced to the eavesdropping attack of Eve in the six-state SARG04-QKD protocol \cite{SARG04:2004:QKD,Tamaki:2006:Unconditionally} given that the two-photon source is used.
With the entanglement distillation techniques \cite{Tamaki:2006:Unconditionally,RevModPhys:09:Scarani}, we can obtain the upper bound of Bob's information $I_{BC}$ about Charlie's conclusive-result bits as (see Appendix B)
\begin{equation}
\begin{aligned}
I_{BC}=H(e_{p}|e_{b}),~~
e_{p}=\frac{2-\sqrt{2}}{4}+\frac{3}{2\sqrt{2}}e_{b},
\end{aligned}
\end{equation}
where $H(e_{p}|e_{b})$ is the conditional Shannon entropy function; $e_{p}$ and $e_{b}$ are the phase and bit error rates, respectively. If a bit is flipped, the probability of a phase shift will be $\frac{4+\sqrt{2}}{8}$. In the QDS protocol, $e_{b}$ is the expectation value of mismatching rate between Alice's bits and Charlie's conclusive results in the untested portion.
Therefore $e_{b}=e_{C}^{c}+\delta_{1}$, $\delta_{1}$ is the finite sample size effect which can be quantified by the random sampling without replacement \cite{korzh:2015:provably} with the failure probability $\epsilon_{1}$. In order to optimally implement the forgery attack, Bob will attempt to make the mismatching rate between his guessed bits and Charlie's conclusive-result bits reach the minimum value $S_{c}$. Employing the properties of min-entropy and max-entropy \cite{renner:2005:security,tomamichel:2012:tight},
$S_{c}$ can be given by
\begin{equation} \label{eq2}
\begin{aligned}
H(S_{c})\geq 1-I_{BC},
\end{aligned}
\end{equation}
where $H(x)=-x\log_{2}x-(1-x)\log_{2}(1-x)$ is the binary Shannon entropy function. There is no chance for Bob to make a successful forgery in the ideal case when Charlie's verification threshold  satisfies $T_{v}<S_{c}$. However, considering the sampling with finite number of independent Bernoulli random values, the observed average value can be less than the expectation value with a probability quantified by the Chernoff bound \cite{chernoff:1952:measure,curty:2014:finite}. Thus, the probability that Charlie accepts (CA) the message forged by Bob is negligible as
\begin{equation} \label{PCA}
\begin{aligned}
\varepsilon_{1}=\textrm{Pr}(CA)\leq \exp[-\frac{(S_{c}-T_{v})^2}{2S_{c}}P_{C}^{c}M_{u}],
\end{aligned}
\end{equation}
where $P_{C}^{c}M_{u}$ is the number of Charlie's untested conclusive-result bits.

When Alice repudiates, Bob and Charlie are automatically regarded as honest.
A successful repudiation happens when Alice disavows the signature, with the message accepted by Bob and rejected by Charlie.
Alice must treat each quantum state received by both Bob and Charlie in the same way since the conclusive results of quantum states they acquire are random, which is similar to Ref. \cite{Dunjko:2014:Quantum}. Conclusive results of partial quantum states can be acquired by Bob and Charlie simultaneously, which can be used in the estimation stage to estimate the correlation strength between the quantum states they receive. Thus, exploiting the technique introduced in Ref.~\cite{Dunjko:2014:Quantum}, our QDS scheme can guarantee security against the coherent repudiation attack even if the symmetry operation (optical multiport) \cite{Dunjko:2014:Quantum} is removed. Let $\Delta_{t}$ ($\Delta$) be the mismatching rate between Bob's and Charlie's test (untested) bit string when they both have conclusive results.
Let $P_{B}$ ($P_{C}$) be the expectation value of the mismatching rate between Alice's and Bob's (Charlie's) untested bits. Due to random sampling, we have $\Delta_{t}\leq e_{B}^{c}+e_{C}^{c}$, $\Delta_{t}+\delta_{2}=\Delta\geq P_{C}/P_{C}^{c}-P_{B}/P_{B}^{c}$, where $\delta_{2}$ is the finite sample size effect quantified by the random sampling without replacement  \cite{korzh:2015:provably} with failure probability $\epsilon_{2}$.
When the two thresholds satisfy $T_{v}>T_{a}+\Delta$, the probability that Bob accepts (BA) a message and Charlie rejects (CR) it is negligible as
\begin{equation} \label{Prep}
\begin{aligned}
\varepsilon_{2}=\textrm{Pr}(BA, CR)=\exp\left[-\frac{(A-P_{B}^{c}T_{a})^2}{2A}M_{u}\right],
\end{aligned}
\end{equation}
where $A=P_{B}$ is a  physical solution of the following equation and inequalities
\begin{equation} \label{sol}
\begin{aligned}
\frac{(A-P_{B}^{c}T_{a})^2}{2A}&=\frac{\left[P_{C}^{c}T_{v}-P_{C}^{c}\left(A/P_{B}^{c}+\Delta\right)\right]^2}{3P_{C}^{c}\left(A/P_{B}^{c}+\Delta\right)},\\
P_{B}^{c}T_{a}<A&<P_{B}^{c}(T_{v}-\Delta).
\end{aligned}
\end{equation}

The robustness quantifies the probability that Bob rejects (BR) a message with the absence of an adversary. The probability can be given by
\begin{equation} \label{rob}
\begin{aligned}
\varepsilon_{\textrm{rob}}=\textrm{Pr}(BR)<&h[P_{B}^{c}M_{u},P_{B}^{c}M_{t},e_{B}^{c}, T_{a}-e_{B}^{c}],\\
h(n,k,\lambda,t)=&\frac{\exp[-\frac{nkt^{2}}{2(n+k)\lambda(1-\lambda)}]C(n,k,\lambda)}{\sqrt{2\pi nk\lambda(1-\lambda)/(n+k)}},\\
C(n,k,\lambda)=\textrm{exp}\Big(&\frac{1}{8(n+k)}+\frac{1}{12k}-\frac{1}{12k\lambda+1}\\
&-\frac{1}{12k(1-\lambda)+1}\Big).
\end{aligned}
\end{equation}
Detailed security analysis and calculation can be founded in Appendix B.

\section{Decoy-state QDS}

The above idea using two ideal single-photon sources can be practically implemented by weak coherent states with the decoy-state method \cite{wang:2005:Beating,Lo:2005:Decoy}.
The schematic layout of our practical QDS with phase-randomized weak coherent states, Scheme I and II, are shown in Fig. 1.
\begin{figure}[tbh]
\centering \resizebox{8cm}{!}{\includegraphics{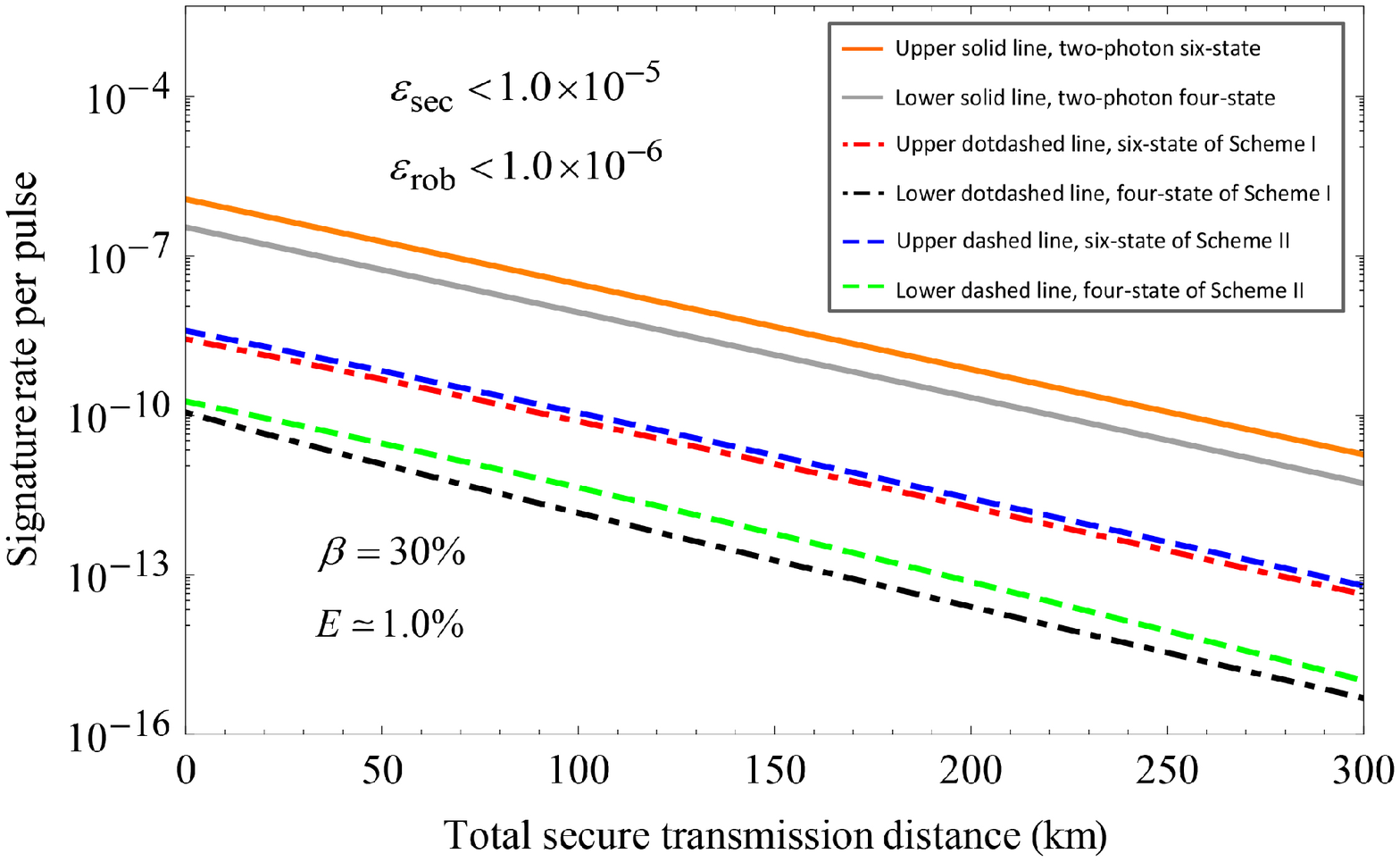}} \caption{(color online). Signature rates versus total secure transmission distance.
For simulation purposes, we employ the following experimental parameters: the intrinsic loss coefficient of the ultralow-loss telecom fiber channel is 0.160 $\textrm{dB}~\textrm{km}^{-1}$ \cite{korzh:2015:provably}. For superconducting nanowire single-photon detectors, the detection efficiency is 93\% and the dark count rate is $1.0\times10^{-7}$ \cite{Marsili:2013:Detector}.
The quantum bit error rate between Alice's and Bob's (Charlie's) system is $E\simeq1.0\%$.
The ratio of random sampling is $\beta=M_{t}/M=30\%$. The security (robustness) bound is $\varepsilon_{\textrm{sec}}<1.0\times10^{-5}$ ($\varepsilon_{\textrm{rob}}<1.0\times10^{-6}$).
The verification and authentication thresholds $\{T_{v},T_{a}\}$ of two-photon six-state (four-state) QDS is $\{6.45\%,1.50\%\}$ ($\{4.05\%,1.20\%\}$); $\{T_{v},T_{a}\}$ of six-state and four-state in Scheme I (II) are $\{4.07\%,1.17\%\}$ ($\{4.275\%,1.20\%\}$) and $\{3.46\%,1.12\%\}$ ($\{3.44\%,1.12\%\}$), respectively. The intensities of six-state for Scheme I are $\mu=0.34$, $\nu=0.16$, $\omega=0.01$ and 0. The probabilities of six-state for Scheme I are $P_{\mu}=55\%$, $P_{\nu}=25\%$, $P_{\omega}=18\%$ and $P_{0}=2\%$.
The intensities of four-state for Scheme I are $\mu=0.12$, $\nu=0.08$, $\omega=0.008$ and 0. The probabilities of four-state for Scheme I are $P_{\mu}=52\%$, $P_{\nu}=23\%$, $P_{\omega}=23\%$ and $P_{0}=2\%$.
The intensities of six-state for Scheme II are $\mu_{1}=0.17$, $\nu_{1}=0.08$ and 0. The probabilities of six-state for Scheme II are $P_{\mu_{1}\mu_{1}}=57\%$, $P_{0,\mu_{1}}=1\%$, $P_{\mu_{1},0}=1\%$, $P_{\nu_{1},\nu_{1}}=30\%$, $P_{0,\nu_{1}}=5\%$, $P_{\nu_{1},0}=5\%$ and $P_{00}=1\%$.
The intensities of four-state for Scheme II are $\mu_{1}=0.075$, $\nu_{1}=0.04$ and 0. The probabilities of four-state for Scheme II are $P_{\mu_{1}\mu_{1}}=60\%$, $P_{0,\mu_{1}}=1\%$, $P_{\mu_{1},0}=1\%$, $P_{\nu_{1},\nu_{1}}=27\%$, $P_{0,\nu_{1}}=5\%$, $P_{\nu_{1},0}=5\%$ and $P_{00}=1\%$.
}
\label{f2}
\end{figure}
Scheme I and Scheme II have a few additional parts than the above two-photon six-state QDS scheme, i.e., decoy-state modulation and announcement.
In the distribution stage of Scheme I, Alice exploits amplitude modulator (AM) to randomly prepare weak coherent state pulses with four intensities,  $\mu,~\nu,~\omega,~0$ ($\mu>\nu>\omega>0$). Their probability distributions are set as $P_{\mu}$, $P_{\nu}$, $P_{\omega}$ and $P_{0}$. Two copies of quantum states can be generated by 50:50 beam splitter (BS).
The phases of Alice's signal laser pulses can be internally modulated \cite{YLT:2014:MDI}, which guarantees the security against unambiguous-state-discrimination attack \cite{Tang:2013:USD}.  In the distribution stage of Scheme II, with a 50:50 BS and two AMs, two copies of weak coherent state pulses are generated and utilized to modulate the following seven sets of intensities: $\{\mu_{1},\mu_{1}\}$, $\{\mu_{1}, 0\}$, $\{0, \mu_{1}\}$, $\{\nu_{1}, \nu_{1}\}$, $\{\nu_{1}, 0\}$, $\{0, \nu_{1}\}$ and $\{0, 0\}$ with $\mu_{1}>\nu_{1}>0$.  Their probability distributions are set as $P_{\mu_{1}\mu_{1}}$, $P_{\mu_{1}0}$, $P_{0\mu_{1}}$, $P_{\nu_{1}\nu_{1}}$, $P_{\nu_{1}0}$, $P_{0\nu_{1}}$ and $P_{00}$.   We define $\{\mu_{1}, \mu_{1}\}$ as the signal-state set, while other six sets compose the decoy-state set.
In the estimation stage of Scheme I (II), Alice announces the intensity information of each pulse, the polarization information of all decoy states (decoy-state sets), and a portion $\beta$ of the signal state (signal-state set) that is randomly selected by Bob or Charlie. For the signal state (signal-state set), the $\beta$ portion of data are regarded as test bits, while the remaining ones are untested bits to be used in the messaging stage.

The randomly selected test bits are used to defeat the repudiation attack.
The decoy-state method is used to estimate the yield and bit error rate of two-photon component, which are exploited to defeat the forgery attack.
For Alice's repudiation attack, the decoy-state method does not bring any advantage because the decoy state data are not used for test bits.
For Bob's forgery attack, the decoy-state method cannot bring any advantage since the decoy states are randomly prepared by Alice. The standard error analysis method \cite{Ma:2005:QKD} is used for estimating the statistical fluctuation in the decoy-state method.
We remark that the roles of Bob and Charlie in Scheme I (II) are equivalent, either can be the authenticator.
Detailed analysis and calculations are shown in Appendix C.

We simulate the signature rates $R$ of our QDS schemes as functions of total secure transmission distance $L$, as shown in Fig. 2.
Here, the signature rate is defined as $R=1/(2N)$ because we employ $2N$ qubit states to sign one bit of classical message.
We consider the symmetric case where the distance between Alice and Bob (Charlie) is $L/2$.
For weak coherent states in Schemes I and II, we present an analytical method to estimate the parameters of two-photon component. Figure 2 also shows signature rates of four-state QDS for comparison. The linear optical elements and threshold single-photon detectors constituting measurement devices are used in our QDS schemes by the universal squash model \cite{fung:2011:threshold}.
The detector error model \cite{Fred:2007:Performance} is applied to calculate the detection probability and error rate of quantum states.
According to the simulation result, if we consider the system with $\textrm{10 GHz}$ clock rate and six-state polarization encoding, we can generate a signature rate of $\textrm{294 bps}$ for a channel length of $\textrm{100km}$ with two copies of single-photon source, $\textrm{0.78 bps}$ ($\textrm{1.12 bps}$) for Scheme I (II) with phase-randomized weak coherent states over the same distance.

\section{Conclusion}

In this work, we propose a QDS protocol with the immediate feasibility of implementing it over a distance of more than 100 km. The signature rate in our protocol can achieve better performance at longer distance than previous protocols due to, among others, the Chernoff bound and the decoy-state method.
We anticipate that the signature rate could be significantly increased by adopting tighter bound in the sampling theory, e.g., the tighter Chernoff bound in \cite{YLT:2014:MDI}.
Similar to QKD, any photon-number distribution source, such as the coherent-state superpositions \cite{yin:2014:long}, can be used for QDS.
It may not be easily generalized to more participants with the scheme in this paper. In QDS with more participants, other aspects should be considered, such as colluding attack \cite{Arrazola:2015:Multi}, efficiency and resource, which should be studied in the future.
We remark that after submitting our manuscript, we became aware of another independent work implementing QDS with unauthenticated quantum channels, however, it still requires secure classical channels \cite{amiri:2015:secure}.

\acknowledgments
This work is supported by the Chinese Academy of Sciences and the National Natural Science Foundation of China under Grant No. 61125502.

H.-L.Y. and Y.F. contributed equally to this work.

\appendix

\section{DETAILED SECURITY ANALYSIS}

\subsection{Security against repudiation}

In a repudiation case, the dishonest signer Alice successfully cheats two honest recipients Bob and Charlie.
Here, Charlie selects the test bits which can be regarded as the random sampling since Charlie is an honest participant.
We exploit the technique introduced in Ref.~\cite{Dunjko:2014:Quantum} to prove that the general quantum repudiation attack (i.e., coherent attack) relative to individual repudiation attack does not provide any advantage.
For each possible future message to be signed, $m=0$ and $m=1$, Alice sends $N$ quantum states $\rho_{B_{1},C_{1},\ldots,B_{N}C_{N}}$ (arbitrary form) to Bob and Charlie, respectively. Therein, $M$ quantum states are received both by Bob and Charlie. Bob and Charlie directly measure the received quantum states and store them as classical bits.
Charlie randomly selects $M_{t}$ bits from $M$ bits as the test bits in the estimation stage. The remaining $M_{u}=M-M_{t}$ untested bits are used in the messaging stage.
If and only if both Bob and Charlie receive a quantum state can the event be used for providing the security against repudiation. Otherwise, Alice can simply make Bob accept the message but Charlie reject it.
In the six-state QDS, for each received quantum state, the probability is $P^{c}=1/6$ that Bob (Charlie) has a conclusive result combining with the set of quantum state in the ideal case ($P^{c}=1/4$ for four-state QDS). Alice is not able to know which quantum state can be confirmed by Bob or Charlie. Therefore, from the perspective of Alice, she must treat each quantum state received by Bob (Charlie) in the same way.
We define $1(0)$ as the event that the measurement result of Bob or Charlie mismatches (matches) Alice's announcement. If Bob or Charlie does not have a conclusive result, the result is considered to match Alice's announcement automatically. We encode each matching result of Bob's and Charlie's untested bits as classical-quantum state $\ket{\varphi_{B}^{u}}$ and $\ket{\varphi_{C}^{u}}$, respectively, with $\varphi_{B}^{u}, \varphi_{C}^{u}=0, 1$.

The classical-quantum state of matching outcomes of Bob and Charlie used in the messaging stage can be written as
\begin{equation} \label{MBC}
\begin{aligned}
\rho_{B}^{u}&=\sum_{\varphi_{B_{1}}^{u},\ldots,\varphi_{B_{M_{u}}}^{u}}p(\varphi_{B_{1}}^{u},\ldots,\varphi_{B_{M_{u}}}^{u})\bigotimes_{i=1}^{M_{u}}\ket{\varphi_{B_{i}}^{u}}\bra{\varphi_{B_{i}}^{u}},\\
\rho_{C}^{u}&=\sum_{\varphi_{C_{1}}^{u},\ldots,\varphi_{C_{M_{u}}}^{u}}p(\varphi_{C_{1}}^{u},\ldots,\varphi_{C_{M_{u}}}^{u})\bigotimes_{i=1}^{M_{u}}\ket{\varphi_{C_{i}}^{u}}\bra{\varphi_{C_{i}}^{u}}.
\end{aligned}
\end{equation}
The successful repudiation probability relies on the state $\rho^{u}$,
\begin{equation} \label{re}
\begin{aligned}
\rho^{u}=\rho_{B}^{u}\otimes\rho_{C}^{u}.
\end{aligned}
\end{equation}
The repudiation process can be described by the following classical-quantum state
\begin{equation} \label{r1}
\begin{aligned}
\mathcal{R}ep(\rho^{u})=p(\rho^{u})\ket{Suc}\bra{Suc}+[1-p(\rho^{u})]\ket{Fai}\bra{Fai},
\end{aligned}
\end{equation}
where the orthogonal states \ket{Suc} and \ket{Fai} represent success and failure of repudiation, respectively. The state $\rho^{u}$ is a convex combination of states,
\begin{equation} \label{r2}
\begin{aligned}
\rho^{u}&=\sum_{\varphi_{B_{1}}^{u},\varphi_{C_{1}}^{u},\ldots,\varphi_{B_{M_{u}}}^{u},\varphi_{C_{M_{u}}}^{u}}p(\varphi_{B_{1}}^{u},\ldots,\varphi_{B_{M_{u}}}^{u})\\
&p(\varphi_{C_{1}}^{u},\ldots,\varphi_{C_{M_{u}}}^{u})\bigotimes_{i=1}^{M_{u}}\ket{\varphi_{B_{i}}^{u}}\bra{\varphi_{B_{i}}^{u}}\ket{\varphi_{C_{i}}^{u}}\bra{\varphi_{C_{i}}^{u}}\\
&=\sum_{j}p(j)\rho_{j}.
\end{aligned}
\end{equation}
The successful repudiation probability is given by
\begin{equation} \label{r3}
\begin{aligned}
\textrm{Tr}&[\ket{Suc}\bra{Suc}\mathcal{R}ep(\rho^{u})]\\
&=\textrm{Tr}\left[\ket{Suc}\bra{Suc}\mathcal{R}ep\left(\sum_{j}p(j)\rho_{j}\right)\right]\\
&=\sum_{j}p(j)\textrm{Tr}[\ket{Suc}\bra{Suc}\mathcal{R}ep(\rho_{j})]\\
&=\sum_{j}p(j)p(\rho_{j}).
\end{aligned}
\end{equation}
Since $\sum_{j}p(j)p(\rho_{j})$ is a convex combination of probabilities and $\sum_{j}p(j)=1$, we have
\begin{equation} \label{r4}
\begin{aligned}
\sum_{j}p(j)p(\rho_{j})\leq \sum_{j}p(j)\max_{j}p(\rho_{j})=\max_{j}p(\rho_{j}).
\end{aligned}
\end{equation}
Therefore, the maximum value of the successful repudiation probability is acquired based on one individual state $\rho_{j}$. That is, the optimal individual repudiation attack can give out the upper bound of the repudiation attack.

In the individual repudiation attack, Alice sends individual and possibly different quantum states to Bob and Charlie. Each pair of the quantum states Bob and Charlie receive are not correlated with others and not required to be identical.
The matching result of Bob's and Charlie's untested bits (used in the messaging stage) can be regarded as independent Bernoulli random variables that satisfy $\textrm{Pr}(\varphi_{B_{i}}^{u}=1)=p_{B}^{i}$ and $\textrm{Pr}(\varphi_{C_{i}}^{u}=1)=p_{C}^{i}$, $i\in\{1,2,\ldots,M_{u}\}$. Let $\bar{X}_{B}=1/M_{u}\sum_{i}\varphi_{B_{i}}^{u}$ and $\bar{X}_{C}=1/M_{u}\sum_{i}\varphi_{C_{i}}^{u}$,
the expectation values of $\bar{X}_{B}$ and $\bar{X}_{C}$ are denoted as $P_{B}=1/M_{u}\sum_{i}p_{B}^{i}$ and $P_{C}=1/M_{u}\sum_{i}p_{C}^{i}$, respectively. An observed outcome of $\bar{X}_{B}$ ($\bar{X}_{C}$) is represented as $\bar{x}_{B}$ ($\bar{x}_{C}$).

If the authentication mismatching rate satisfies $\bar{x}_{B}\leq P_{B}^{c}T_{a}$ ($P_{B}^{c}$ represents the probability of Bob's conclusive results), Bob accepts the signed message. By exploiting Chernoff Bound \cite{chernoff:1952:measure,curty:2014:finite}, the probability of Bob accepting (denoted as BA) a valid message can be given by
\begin{equation} \label{PBA}
\begin{aligned}
\textrm{Pr}(BA)&=\textrm{Pr}(P_{B}-\bar{x}_{B}\geq P_{B}-P_{B}^{c}T_{a}) \\
&\leq\exp\left[-\frac{(P_{B}-P_{B}^{c}T_{a})^2}{2P_{B}}M_{u}\right],
\end{aligned}
\end{equation}
where $\textrm{Pr}(BA)$ is a strictly decreasing function for parameter $P_{B}$, $1\geq P_{B}\geq P_{B}^{c}T_{a}$.
If the verification mismatching rate $\bar{x}_{C}\geq P_{C}^{c}T_{v}$ ($P_{C}^{c}$ represents the probability of Charlie's conclusive results), Charlie will reject the signed message. By exploiting Chernoff Bound \cite{chernoff:1952:measure,curty:2014:finite}, the probability of Charlie rejecting (denoted as CR) a valid message can be given by
\begin{equation} \label{PCV}
\begin{aligned}
\textrm{Pr}(CR)&=\textrm{Pr}(\bar{x}_{C}-P_{C}\geq P_{C}^{c}T_{v}-P_{C})\\
&\leq \exp\left[-\frac{(P_{C}^{c}T_{v}-P_{C})^2}{3P_{C}}M_{u}\right],
\end{aligned}
\end{equation}
where $\textrm{Pr}(BA)$ is a strictly increasing function for parameter $P_{C}$, $0<P_{C}<P_{C}^{c}T_{v}$.

A successful repudiation means that Bob accepts the signed message and Charlie rejects it. The probability can be written as
\begin{equation} \label{Prep}
\begin{aligned}
\textrm{Pr}(BA,CR)\leq \sup\{\min\{\textrm{Pr}(BA),\textrm{Pr}(CR)\}\}.
\end{aligned}
\end{equation}
Under reasonable conditions, in order to make the value of $\textrm{Pr}(BA, CR)$ as large as possible, Alice will make the parameter $P_{C}$ as large as possible and $P_{B}$ as small as possible.
Because we remove the SWAP test \cite{Gottesman:2001} and symmetry operation (optical multiport) \cite{Dunjko:2014:Quantum} (to guarantee that the quantum states that Bob and Charlie receive are identical),  $P_{B}=P_{C}$ will no longer be satisfied. However, we can restrict the difference between $P_{C}$ and $P_{B}$. Thereby, the upper bound of the successful repudiation probability can be restricted.

\begin{figure}[tbh]
\centering \resizebox{6cm}{!}{\includegraphics{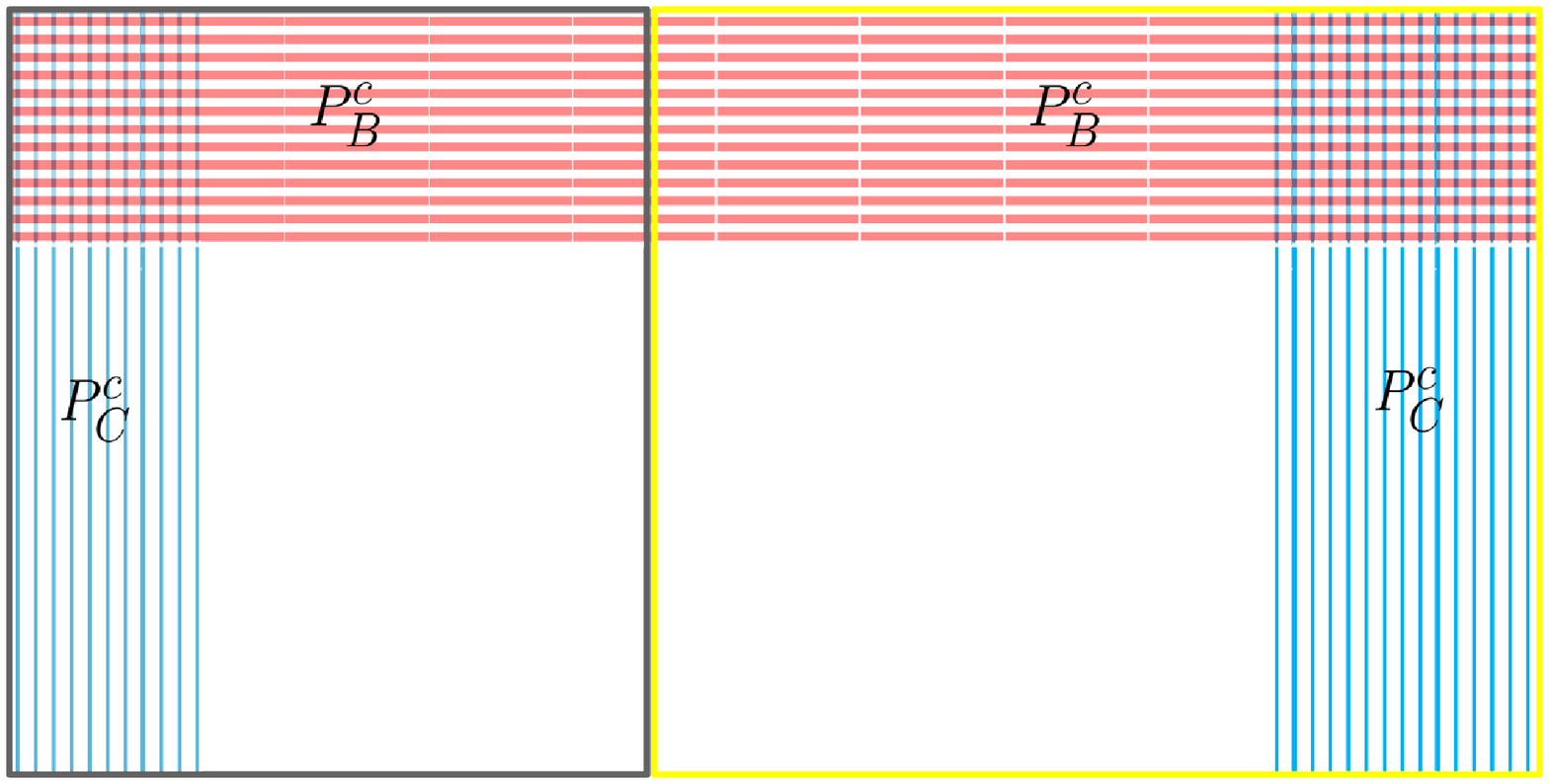}} \caption{(color online). Schematic representation of different bit strings.
The left black frame (right yellow frame) rectangle represents the string of test (untested) bits. The horizontal red shadow (vertical blue shadow) represents the bit string of Bob's (Charlie's) conclusive results.}
\label{f3}
\end{figure}

For all quantum states received by both Bob and Charlie, Bob (Charlie) records a string of data $\mathcal{Y}_{B}=\{y_{B_{1}}, y_{B_{2}}, \ldots, y_{B_{M}}\}$ ($\mathcal{Y}_{C}=\{y_{C_{1}}, y_{C_{2}}, \ldots, y_{C_{M}}\}$).
Here, $y_{B_{i}}$ represents the $i$th data and one has $y_{B_{i}}\in\{0, 1, \perp\}$. $y_{B_{i}}=0, 1$ ($y_{B_{i}}=\perp$) represents that Bob has a conclusive (inconclusive) outcome.
Let $\mathcal{Y}_{B}^{t}$ ($\mathcal{Y}_{C}^{t}$) be the test bit string which is a random sample of size $M_{t}$ of $\mathcal{Y}_{B}$ ($\mathcal{Y}_{C}$) and the remaining untested bit string is $\mathcal{Y}_{B}^{u}$ ($\mathcal{Y}_{C}^{u}$).
Similarly, the bit string of Alice is $\mathcal{Y}_{A}$ ($\mathcal{Y}_{A}=\mathcal{Y}_{A}^{t}\bigcup\mathcal{Y}_{A}^{u}$) and we have $y_{A_{i}}\in\{0, 1\}$ for the quantum states prepared by Alice.
As is clear from the above descriptions, we have $\varphi_{B_{i}}^{u}=y_{A_{i}}^{u}\oplus y_{B_{i}}^{u}$ and $\varphi_{C_{i}}^{u}=y_{A_{i}}^{u}\oplus y_{C_{i}}^{u}$ ($0, 1~\oplus \perp:=0$). Let $\mathcal{Y}_{B}^{c}=\mathcal{Y}_{B}^{ct}\cup\mathcal{Y}_{B}^{cu}$ and $\mathcal{Y}_{C}^{c}=\mathcal{Y}_{C}^{ct}\cup\mathcal{Y}_{C}^{cu}$ represent the bit strings of Bob's and Charlie's conclusive results, respectively. $\mathcal{Y}_{AB}^{c}=\mathcal{Y}_{AB}^{ct}\cup\mathcal{Y}_{AB}^{cu}$ represents Alice announcing the bit string corresponding to $\mathcal{Y}_{B}^{c}=\mathcal{Y}_{B}^{ct}\cup\mathcal{Y}_{B}^{cu}$, while $\mathcal{Y}_{AC}^{c}=\mathcal{Y}_{AC}^{ct}\cup\mathcal{Y}_{AC}^{cu}$ represents Alice announcing the bit string corresponding to $\mathcal{Y}_{C}^{c}=\mathcal{Y}_{C}^{ct}\cup\mathcal{Y}_{C}^{cu}$. Let $\mathcal{Z}_{B}^{c}=\mathcal{Z}_{B}^{ct}\cup\mathcal{Z}_{B}^{cu}$ ($\mathcal{Z}_{C}^{c}=\mathcal{Z}_{C}^{ct}\cup\mathcal{Z}_{C}^{cu}$) represents the bit string of Bob (Charlie) given that Bob and Charlie both have conclusive results, and the corresponding bit string of Alice is $\mathcal{Z}_{A}^{c}=\mathcal{Z}_{A}^{ct}\cup\mathcal{Z}_{A}^{cu}$. A visualized schematic of the relationship between the above bit strings are shown in Fig.~\ref{f3}.

Exploiting relative Hamming distance and the random sampling theorem, for the arbitrary bit string Alice announces, the expectation value $\Delta$ can be given by
\begin{equation} \label{dis}
\begin{aligned}
\Delta&=\textrm{d}_{\textrm{H}}(\mathcal{Z}_{B}^{cu}, \mathcal{Z}_{C}^{cu})\geq \textrm{d}_{\textrm{H}}(\mathcal{Z}_{A}^{cu}, \mathcal{Z}_{C}^{cu})-\textrm{d}_{\textrm{H}}(\mathcal{Z}_{A}^{cu}, \mathcal{Z}_{B}^{cu})\\
&=\textrm{d}_{\textrm{H}}(\mathcal{Y}_{AC}^{cu}, \mathcal{Y}_{C}^{cu})-\textrm{d}_{\textrm{H}}(\mathcal{Y}_{AB}^{cu}, \mathcal{Y}_{B}^{cu})\\
&=\textrm{d}_{\textrm{H}}(\mathcal{Y}_{A}^{u}, \mathcal{Y}_{C}^{u})/P_{C}^{c}-\textrm{d}_{\textrm{H}}(\mathcal{Y}_{A}^{u}, \mathcal{Y}_{B}^{u})/P_{B}^{c}\\
&=P_{C}/P_{C}^{c}-P_{B}/P_{B}^{c}.\\
\end{aligned}
\end{equation}
The test distance $\Delta_{t}$ can be given by
\begin{equation} \label{d1}
\begin{aligned}
\Delta_{t}&=\textrm{d}_{\textrm{H}}(\mathcal{Z}_{B}^{ct}, \mathcal{Z}_{C}^{ct})\leq \textrm{d}_{\textrm{H}}(\mathcal{Z}_{A}^{ct}, \mathcal{Z}_{B}^{ct})+\textrm{d}_{\textrm{H}}(\mathcal{Z}_{A}^{ct}, \mathcal{Z}_{C}^{ct})\\
&= \textrm{d}_{\textrm{H}}(\mathcal{Y}_{AB}^{ct}, \mathcal{Y}_{B}^{ct})+\textrm{d}_{\textrm{H}}(\mathcal{Y}_{AC}^{ct}, \mathcal{Y}_{C}^{ct})
=e_{B}^{c}+e_{C}^{c}.
\end{aligned}
\end{equation}
Here, $e_{B}^{c}$ ($e_{C}^{c}$) is the mismatching rate between Bob's (Charlie's) conclusive results of test bits and Alice's announcement bits, which can be acquired in the estimation stage.
Taking into account the random sampling without replacement theorem \cite{korzh:2015:provably}, we have
\begin{equation} \label{d2}
\begin{aligned}
\Delta=\Delta_{t}+\delta_{2},~~~
\delta_{2}=&g[P_{C}^{c}P_{B}^{c}M_{u},P_{C}^{c}P_{B}^{c}M_{t},\Delta_{t},\epsilon_{2}],\\
\end{aligned}
\end{equation}
\begin{equation}
\begin{aligned}
g(n,k,\lambda,\bar{\epsilon})=&\sqrt{\frac{2(n+k)\lambda(1-\lambda)}{nk}}\\
&\times\sqrt{\ln\frac{\sqrt{n+k}C(n,k,\lambda)}{\sqrt{2\pi nk\lambda(1-\lambda)}\bar{\epsilon}}},\\
C(n,k,\lambda)=&\textrm{exp}\Big(\frac{1}{8(n+k)}+\frac{1}{12k}-\frac{1}{12k\lambda+1}\\
&-\frac{1}{12k(1-\lambda)+1}\Big).
\end{aligned}
\end{equation}
where $\epsilon_{2}$ is the failure probability, $P_{C}^{c}P_{B}^{c}M_{u}$ and $P_{C}^{c}P_{B}^{c}M_{t}$ are the numbers of untested bits and test bits given that both Bob and Charlie have the conclusive results, respectively.

From Eq.~\eqref{PBA}-Eq.~\eqref{Prep}, we know that the optimal probability that Bob accepts a message while Charlie rejects it is
\begin{equation} \label{opt}
\begin{aligned}
\varepsilon_{2}=\textrm{Pr}(BA, CR)=\exp\left[-\frac{(A-P_{B}^{c}T_{a})^2}{2A}M_{u}\right],
\end{aligned}
\end{equation}
where $A=P_{B}$ is a  physical solution of the following equation and inequalities
\begin{equation} \label{sol}
\begin{aligned}
\frac{(A-P_{B}^{c}T_{a})^2}{2A}&=\frac{\left[P_{C}^{c}T_{v}-P_{C}^{c}\left(A/P_{B}^{c}+\Delta\right)\right]^2}{3P_{C}^{c}\left(A/P_{B}^{c}+\Delta\right)},\\
P_{B}^{c}T_{a}<A&<P_{B}^{c}(T_{v}-\Delta).
\end{aligned}
\end{equation}
The protocol is  with $\varepsilon_{\textrm{nor}}$-non-repudiation, where $\varepsilon_{\textrm{nor}}=\varepsilon_{2}+\epsilon_{2}$.

\subsection{Security against forgery}
In a forgery case, the dishonest recipient, Bob, can successfully deceive two honest participants, i.e., the signer Alice and the other recipient Charlie. Bob authenticates the validity of the message sent by Alice, while Charlie verifies the validity of the message forwarded by Bob.
Here, the test bits Charlie selects can be regarded as the random sampling since Charlie is an honest participant.
The forgery is successful when Charlie accepts (denoted as CA) the forged (tampered) message forwarded by Bob.
For all conclusive measurement outcomes of the untested bits, Charlie records a string of data $\mathcal{Y}_{C}^{cu}=\{y_{C_{1}}^{cu}, y_{C_{2}}^{cu}, \ldots, y_{C_{L_{c}}}^{cu}\}$, where  $L_{c}=P_{C}^{c}M_{u}$ is the number of conclusive measurement outcomes of the untested bits.
It is obvious to see $\mathcal{Y}_{C}^{cu}\subseteq\mathcal{Y}_{C}^{u}$ and $y_{C_{i}}^{cu}\in\{0, 1\}$ for $i\in\{1, \ldots, L_{c}\}$.
Let $\mathcal{Y}_{BF}^{cu}=\{y_{BF_{1}}^{cu}, y_{BF_{2}}^{cu}, \ldots, y_{BF_{L_{c}}}^{cu}\}$ represents the bit string Bob forwards corresponding to $\mathcal{Y}_{C}^{cu}$.
We consider the case that Bob is restricted to collective forgery attack in which his optimal strategy is to correctly guess the information of each quantum state as much as possible.
We assume that the upper bound of Bob's information $I_{BC}$ about Charlie's conclusive result bits can be acquired with failure probability $\epsilon_{1}$.
Note that the values of estimating the information are different for various protocols and will be analyzed in the following section.
Let $S_{c}$ be the mismatching rate between $\mathcal{Y}_{BF}^{cu}$ and $\mathcal{Y}_{C}^{cu}$. In order to optimally implement the forgery attack, Bob will make $S_{c}$ reach the minimum value with $I_{BC}$.
So the minimum mismatching rate $S_{c}$ can be given by \cite{tomamichel:2012:tight,renner:2005:security}
\begin{equation} \label{err}
\begin{aligned}
H(S_{c})\geq H_{max}(C|B)\geq H_{min}(C|B)=1-I_{BC},
\end{aligned}
\end{equation}
where $H(x)=-x\log_{2}x-(1-x)\log_{2}(1-x)$ is the binary entropy function.
The max-entropy $H_{max}(C|B)$ is upper bounded by the minimum number of bits of additional information about $\mathcal{Y}_{C}^{cu}$ which are needed to perfectly reconstruct $\mathcal{Y}_{C}^{cu}$ from $\mathcal{Y}_{BF}^{cu}$.
The min-entropy $H_{min}(C|B)$ quantifies the minimum uncertainty that Bob has about $\mathcal{Y}_{C}^{cu}$ using the optimal attack strategy.

Let $\mathcal{X}_{C}^{c}=\{\varphi_{C_{1}}^{cu}, \varphi_{C_{2}}^{cu}, \ldots, \varphi_{C_{L_{c}}}^{cu}\}$ be a set of independent Bernoulli random variables which represent the matching results between Charlie's conclusive measurement outcomes of the untested bits and those Bob forwards, i.e., $\varphi_{C_{i}}^{cu}=y_{C_{i}}^{cu}\oplus y_{BF_{i}}^{cu}$. Let $\bar{X}_{C}^{c}=1/L_{c}\sum_{i}\varphi_{C_{i}}^{cu}$, the expectation value of $\bar{X}_{C}^{c}$ can be given by $E[\bar{X}_{C}^{c}]=S_{c}$.
An observed outcome of $\bar{X}_{C}^{c}$ is represented as $\bar{x}_{C}^{c}$.
If the verification mismatching rate satisfies $\bar{x}_{C}^{c}\leq T_{v}$, Charlie accepts the forged message.
There is no chance that Bob successfully forges in the ideal conditions when the verification threshold of Charlie satisfies $T_{v}<S_{c}$. However, considering the sampling with the case of finite number of independent Bernoulli random values, the observed average value can be less than the expectation value with a negligible probability.
The probability that Charlie accepts (denoted as CA) a forged message can be given by
\begin{equation} \label{PCA}
\begin{aligned}
\varepsilon_{1}&=\textrm{Pr}(CA)=\textrm{Pr}(S_{c}-\bar{x}_{C}^{c}\geq S_{c}-T_{v})\\
&\leq \exp[-\frac{(S_{c}-T_{v})^2}{2S_{c}}L_{c}]\\
&=\exp[-\frac{(S_{c}-T_{v})^2}{2S_{c}}P_{C}^{c}M_{u}].
\end{aligned}
\end{equation}
The protocol is  with $\varepsilon_{\textrm{unf}}$-unforgeability, where $\varepsilon_{\textrm{unf}}=\varepsilon_{1}+\epsilon_{1}$.
Therefore, the security level of the QDS can be given by
\begin{equation} \label{sol}
\begin{aligned}
\varepsilon_{\textrm{sec}}&=\varepsilon_{\textrm{nor}}+\varepsilon_{\textrm{unf}}\\
&=\varepsilon_{1}+\varepsilon_{2}+\epsilon_{1}+\epsilon_{2}.
\end{aligned}
\end{equation}

\subsection{The robustness}

$\varepsilon_{\textrm{rob}}$ is the probability that the protocol is aborted when the adversary is inactive. In the estimation stage, let $e_{B}^{c}$ be the mismatching rate of conclusive results of Bob's test bits.  Exploiting random sampling without replacement theorem \cite{korzh:2015:provably}, in the messaging stage, the mismatching rate $E_{B}^{c}$ of conclusive results of Bob's untested bits can be given by (with no adversary existing)
\begin{equation} \label{EB}
\begin{aligned}
E_{B}^{c}=e_{B}^{c}+g[P_{B}^{c}M_{u},P_{B}^{c}M_{t},e_{B}^{c},\varepsilon'],
\end{aligned}
\end{equation}
where $\varepsilon'$ is the failure probability, $P_{B}^{c}M_{u}$ and $P_{B}^{c}M_{t}$ are the numbers of untested bits and test bits given that Bob has conclusive results, respectively.
If one has $E_{B}>T_{a}$, Bob rejects (denoted as BR) the message sent by Alice. The probability can be written as
\begin{equation} \label{rob}
\begin{aligned}
\varepsilon_{\textrm{rob}}=\textrm{Pr}(BR)<\varepsilon'=&h(P_{B}^{c}M_{u},P_{B}^{c}M_{t},E_{B}^{c}, T_{a}-E_{B}^{c}),\\
h(n,k,\lambda,t)=&\frac{\exp[-\frac{nkt^{2}}{2(n+k)\lambda(1-\lambda)}]C(n,k,\lambda)}{\sqrt{2\pi nk\lambda(1-\lambda)/(n+k)}},\\
\end{aligned}
\end{equation}

\section{Two copies of single-photon states}

In the following, we analyze Bob's information $I_{BC}$ about Charlie's conclusive result bits in the two-photon six-state (four-state) QDS in detail.
Besides, we calculate the signature rate and the corresponding security bound in the practical fiber-based protocol.

We consider the case that Bob is restricted to collective forgery attack in which his optimal strategy is to correctly guess the information of each quantum state as much as possible.
Therefore, the attack strategy of Bob in two-photon six-state QDS protocol is equivalent to the eavesdropping attack of Eve in the six-state SARG04-QKD protocol \cite{SARG04:2004:QKD,Tamaki:2006:Unconditionally} given that the two-photon source has been taken into account (In the QKD protocol, the two communication parties, Alice and Bob, trust each other while the eavesdropper Eve is the untrusted adversary).
In the unconditionally secure SARG04-QKD protocol \cite{Tamaki:2006:Unconditionally}, an virtual entanglement-based protocol is proposed.
Exploiting the unconditionally secure entanglement distillation protocol with two-photon, the upper bound of $I_{BC}=I_{E}=H(e_{p}|e_{b})$ can be estimated \cite{RevModPhys:09:Scarani}. The relationship between phase error rate $e_{p}$ and bit error rate $e_{b}$ with the six-state SARG04-QKD \cite{Tamaki:2006:Unconditionally} cannot be provided. Therefore, we generalize the method proposed in Ref~\cite{Tamaki:2006:Unconditionally} to find out the relationship.

Some notations should be defined. The four quantum states are written as $\ket{\varphi_{0}}=\cos\frac{\pi}{8}\ket{0_{x}}+\sin\frac{\pi}{8}\ket{1_{x}}$, $\ket{\bar{\varphi}_{0}}=-\sin\frac{\pi}{8}\ket{0_{x}}+\cos\frac{\pi}{8}\ket{1_{x}}$, $\ket{\varphi_{1}}=\cos\frac{\pi}{8}\ket{0_{x}}-\sin\frac{\pi}{8}\ket{1_{x}}$ and $\ket{\bar{\varphi}_{1}}=\sin\frac{\pi}{8}\ket{0_{x}}+\cos\frac{\pi}{8}\ket{1_{x}}$. Therein, $\ket{\bar{\varphi}_{0}}$ and $\ket{\varphi_{0}}$ are eigenstates of basis $\frac{Z+X}{\sqrt{2}}$, $\ket{\bar{\varphi}_{1}}$ and $\ket{\varphi_{1}}$ are eigenstates of basis $\frac{Z-X}{\sqrt{2}}$. A filtering operator reads $F=\sin\frac{\pi}{8}\ket{0_{x}}\bra{0_{x}}+\cos\frac{\pi}{8}\ket{1_{x}}\bra{1_{x}}$ and a $-\frac{\pi}{2}$ rotation around $Y$ basis reads $R=\cos\frac{\pi}{4}I+\sin\frac{\pi}{4}(\ket{1_{x}}\bra{0_{x}}-\ket{0_{x}}\bra{1_{x}})$, note that $R\ket{\varphi_{1}}=\ket{\varphi_{0}}$. $T_{0}=I$ represents an identity operator, $T_{1}=\cos\frac{\pi}{4}I-i\sin\frac{\pi}{4}\frac{Z+X}{\sqrt{2}}$ represents a $\frac{\pi}{2}$ rotation around $\frac{Z+X}{\sqrt{2}}$ basis, $T_{2}=\cos\frac{\pi}{4}I-i\sin\frac{\pi}{4}\frac{Z-X}{\sqrt{2}}$ represents a $\frac{\pi}{2}$ rotation around $\frac{Z-X}{\sqrt{2}}$ basis.

The entanglement-based protocol can be describe in the following \cite{Tamaki:2006:Unconditionally}. For each quantum signal, Alice prepares an entangled state $\ket{\Psi_{AB}}=(1/\sqrt{2})(\ket{0_{z}}_{A}\ket{\varphi_{0}}_{B}\ket{\varphi_{0}}_{B}+\ket{1_{z}}_{A}\ket{\varphi_{1}}_{B}\ket{\varphi_{1}}_{B})$. Alice randomly applies a rotation $T_{l}R^{k}$ to system B, with $l\in\{0, 1, 2\}$ and $k\in\{0, 1, 2, 3\}$, then she sends system B to Bob through the insecure (unauthenticated) quantum channel. After some possible intervention from Eve, Bob
receives the quantum state. Bob randomly applies the rotation $R^{-k'}T_{l'}^{-1}$ to the qubit state and a filtering operation whose successful operation is descried by Kraus operator $F$. A successful filtering corresponds to a conclusive result of Bob. Alice and Bob then publicly announce $\{k, l\}$ and $\{k', l'\}$, meanwhile, they keep the quantum states with $k=k', l=l'$. Bob randomly chooses some quantum states as test bits. Alice and Charlie measure them in $Z$ basis. Then, they compare the partial measurement outcomes to estimate bit error rates and the information that Eve acquires.

Let $\rho_{\textrm{qubit}}$ represents a pair of qubit states that Alice and Bob share, which can be given by
\begin{equation} \label{qubit}
\begin{aligned}
\rho_{\textrm{qubit}}=\sum_{l,k}\hat{P}[I_{A}\otimes(\sum_{u}FR_{B}^{-k}T_{l_{B}}^{-1}E_{B}^{u}T_{l_{B}}R_{B}^{k})\ket{\xi_{l,k,u}}],
\end{aligned}
\end{equation}
where $l\in\{0,1,2\}$, $k\in\{0,1,2,3\}$, $u\in\{0,1\}$ and
\begin{equation} \label{qubit1}
\begin{aligned}
\ket{\xi_{l,k,u}}=\frac{1}{\sqrt{2}}[&\bra{u_{x}}T_{l}R^{k}\ket{\varphi_{0}}\ket{0_{z}}_{A}\ket{\varphi_{0}}_{B}\\
&+\bra{u_{x}}T_{l}R^{k}\ket{\varphi_{1}}\ket{1_{z}}_{A}\ket{\varphi_{1}}_{B}],\\
E_{B}^{u}=\bra{0_{x}}E_{B}&\ket{u_{x}},~~\hat{P}(X\ket{\Psi})=X\ket{\Psi}\bra{\Psi}X^{\dagger}.
\end{aligned}
\end{equation}
$E_{B}$ is a $4\times4$ matrix which depends on Eve's operation and we can safely assume that the final state of Eve's system is a particular state $\ket{0_{x}}$. The probabilities of bit flip and phase shift can be given by
\begin{equation} \label{probability}
\begin{aligned}
p_{\textrm{bit}}=P_{X}+P_{Y},\\
p_{\textrm{ph}}=P_{Z}+P_{Y},\\
\end{aligned}
\end{equation}
where
\begin{equation} \label{probability1}
\begin{aligned}
P_{X}=\textrm{Tr}[\rho_{\textrm{qubit}}\ket{\Psi^{+}}\bra{\Psi^{+}}],\\
P_{Y}=\textrm{Tr}[\rho_{\textrm{qubit}}\ket{\Psi^{-}}\bra{\Psi^{-}}],\\
P_{Z}=\textrm{Tr}[\rho_{\textrm{qubit}}\ket{\Phi^{-}}\bra{\Phi^{-}}].\\
\end{aligned}
\end{equation}
If $Cp_{\textrm{bit}}+C'p_{\textrm{fil}}\geq p_{\textrm{ph}}$ holds, then $Ce_{b}+C'\geq e_{p}$ is exponentially reliable as the number of successfully filtering states increases \cite{Tamaki:2006:Unconditionally}. $p_{\textrm{fil}}=\textrm{Tr}[\rho_{\textrm{qubit}}]$ is the trace of state $\rho_{\textrm{qubit}}$. It is very clear to see that $p_{\textrm{fil}}$, $p_{\textrm{bit}}$ and $p_{\textrm{ph}}$ are the functions of eight elements and their conjugates, i.e., $p_{\textrm{fil}}=\vec{c}^{*}A_{\textrm{fil}}\vec{c}^{T}$, $p_{\textrm{bit}}=\vec{c}^{*}A_{\textrm{bit}}\vec{c}^{T}$ and $p_{\textrm{ph}}=\vec{c}^{*}A_{\textrm{ph}}\vec{c}^{T}$. $A_{\textrm{fil}}$, $A_{\textrm{bit}}$, and $A_{\textrm{ph}}$ are $8\times 8$ matrices, the eight elements in $\vec{c}$ are directly taken from $E_{B}$. If $CA_{\textrm{bit}}+C'A_{\textrm{fil}}-A_{\textrm{ph}}\geq 0$ is a positive semi-definite matrix, $Cp_{\textrm{bit}}+C'p_{\textrm{fil}}\geq p_{\textrm{ph}}$ will always be satisfied. After a complex calculation according to the above formulas, we can acquire the relationship between phase error rate and bit error rate
\begin{equation} \label{relationship}
\begin{aligned}
e_{p}=\frac{2-\sqrt{2}}{4}+\frac{3}{2\sqrt{2}}e_{b},
\end{aligned}
\end{equation}
in the two-photon six-state SARG04-QKD. Therein, the probability that both bit and phase occur error is $P_{Y}=\frac{4+\sqrt{2}}{8}e_{b}$. With the same method, for the single-photon six-state SARG04-QKD, we have $e_{p}=\frac{3}{2}e_{b}$ and $P_{Y}=\frac{3}{4}e_{b}$.
In the asymptotic case, $I_{BC}$ can be given by
\begin{equation} \label{mutual information}
\begin{aligned}
I_{BC}=&I_{E}=H(e_{p}|e_{b})\\
=&-(1+a-e_{b}-e_{p})\log_{2}\frac{1+a-e_{b}-e_{p}}{1-e_{b}}\\
&-(e_{p}-a)\log_{2}\frac{e_{p}-a}{1-e_{b}}-(e_{b}-a)\log_{2}\frac{e_{b}-a}{e_{b}}\\
&-a\log_{2}\frac{a}{e_{b}},
\end{aligned}
\end{equation}
where $H(e_{p}|e_{b})$ is the conditional Shannon entropy function, $a=P_{Y}=\frac{4+\sqrt{2}}{8}e_{b}$ quantifies the mutual information between bit and phase errors.

In the two-photon four-state QDS scheme, there are four single-photon BB84 quantum states $\ket{H}$, $\ket{V}$, $\ket{+}=(\ket{H}+\ket{V})/\sqrt{2}$, $\ket{-}=(\ket{H}-\ket{V})/\sqrt{2}$. The four states can be divided into four sets $\{\ket{H},\ket{+}\}$, $\{\ket{+},\ket{V}\}$, $\{\ket{V},\ket{-}\}$, $\{\ket{-},\ket{H}\}$, where the first state from each set represents logic 0 and the second state logic 1. In addition to the preparation of quantum states, other processes are the same with the two-photon six-state QDS scheme. The entanglement distillation protocol can be converted to the unconditionally secure two-photon four-state SARG04-QKD. In the asymptotic case, the relationship between the phase error rate and the bit error rate can be given by \cite{Tamaki:2006:Unconditionally}
\begin{equation} \label{relat}
\begin{aligned}
e_{p}=\min_{x}\left\{xe_{b}+\frac{3-2x+\sqrt{6-6\sqrt{2}x+4x^2}}{6}\right\}, \forall x.
\end{aligned}
\end{equation}
Meanwhile, we can set $a=P_{Y}=e_{b}\times e_{p}$, which corresponds to no mutual information between bit and phase errors.

Hereafter, the detector error model \cite{Fred:2007:Performance} is applied to estimate the detection probability and error rate of quantum states. In the simulation, we simply apply the case that Alice and Bob do not interfere with the protocol.
The overall gain $Q$ can be given by
\begin{equation} \label{gain}
\begin{aligned}
Q=[1-(1-Y_{0})(1-\eta_{B})][1-(1-Y_{0})(1-\eta_{C})],
\end{aligned}
\end{equation}
which indicates the ratio of the number of Bob's and Charlie's detection coincidence events to Alice's number of emitted signals. $Y_{0}$ represents the probability that Bob's (Charlie's) detector clicks when the input of Bob (Charlie) is a vacuum state. Because of active basis selection, we have $Y_{0}=2p_{d}(1-p_{d})$, where $p_{d}$ represents the dark count rate of each detector. $\eta_{B}=\eta_{d}\times10^{-\alpha L_{AB}/10}$ ($\eta_{C}=\eta_{d}\times10^{-\alpha L_{AC}/10}$) represents the transmission efficiency from Alice to Bob (Charlie). Here, we consider a widely used fiber-based setup model. Therein, $\eta_{d}$ represents the detection efficiency, $L_{AB}$ ($L_{AC}$) is the distance between Alice and Bob (Charlie), $\alpha$ is the intrinsic loss coefficient of the fiber. Taking into account the universal squash model \cite{fung:2011:threshold}, the threshold single-photon detector can be used in our scheme.
The gain of Bob's conclusive results and Charlie's conclusive results are given by
\begin{equation} \label{gain1}
\begin{aligned}
Q_{B}^{c}&=z[(1-\eta_{B})Y_{0}+(\frac{1}{2}+e_{d})\eta_{B}][1-(1-Y_{0})(1-\eta_{C})]\\
&=P_{B}^{c}Q,\\
Q_{C}^{c}&=z[(1-\eta_{C})Y_{0}+(\frac{1}{2}+e_{d})\eta_{C}][1-(1-Y_{0})(1-\eta_{B})]\\
&=P_{C}^{c}Q,\\
\end{aligned}
\end{equation}
where $e_{d}$ represents the misalignment in the channel, for six-state scheme, $z=\frac{1}{3}$, for four-state scheme, $z=\frac{1}{2}$.  The overall quantum bit error rate (QBER) of Bob's conclusive results and Charlie's conclusive results are given by
\begin{equation} \label{gain1}
\begin{aligned}
e_{B}^{c}Q_{B}^{c}=z[\frac{1}{2}(1-\eta_{B})Y_{0}+e_{d}\eta_{B}][1-(1-Y_{0})(1-\eta_{C})],\\
e_{C}^{c}Q_{C}^{c}=z[\frac{1}{2}(1-\eta_{C})Y_{0}+e_{d}\eta_{C}][1-(1-Y_{0})(1-\eta_{B})],\\
\end{aligned}
\end{equation}
The amount of Charlie's randomly selected test bits is $M_{t}=\beta QN$, the remaining bits $M_{u}=(1-\beta)QN$ are untested bits.
In the QDS protocol, $e_{b}$ is the expectation value of mismatching rate between Alice's bits and Charlie's conclusive result bits in the untested portion.
Therefore $e_{b}=e_{C}^{c}+\delta_{1}$, $\delta_{1}$ is the finite sample size effect which can be quantified by the random sampling without replacement theorem \cite{korzh:2015:provably} with failure probability $\epsilon_{1}$.
Therefore, we have
\begin{equation} \label{inf}
\begin{aligned}
&\delta_{1}=g[P_{C}^{c}M_{u},P_{C}^{c}M_{t},e_{C}^{c},\epsilon_{1}].
\end{aligned}
\end{equation}

Note that the security thresholds satisfy $T_{a}<T_{v}<S_{c}$, the supremum of $S_{c}$ can be given by
\begin{equation} \label{Sc}
\begin{aligned}
H(\sup\{S_{c}\})&=1-H(\inf\{e_{p}|e_{b}\}),\\
&=1-H(\frac{2-\sqrt{2}}{4}).
\end{aligned}
\end{equation}
It is obvious to see that the supremum of $S_{c}$ is 7.9135\% in the two-photon four-state QDS, which is equal to that in the two-photon six-state QDS. However, the six-state scheme is more robust than the one with four-state, for instance, given $e_{b}=1\%$, one has $S_{c}=7.4564\%$ for six-state and $S_{c}=4.5035\%$ for four-state.

\section{Weak coherent states}

\subsection{Practical QDS with Scheme I}
Note that in the case of repudiation, the dishonest signer Alice successfully cheats two honest recipients Bob and Charlie.
Thus, the test bits Charlie or Bob select can be regarded as the random sampling since Charlie (Bob) is honest in the repudiation case.
Therefore the test bits can be used for the security against the repudiation attack.
In the case of forgery, only one recipient is dishonest while the signer is honest. The decoy states are randomly prepared by Alice which can be used to prove the security against forgery attack.
Therefore, the roles of Bob and Charlie are equivalent in the decoy-state-based QDS protocol, both Bob and Charlie could be the authenticator.

The overall gain $Q_{\lambda}$ can be given by
\begin{equation} \label{gain}
\begin{aligned}
Q_{\lambda}=[1-(1-Y_{0})e^{-\frac{\lambda}{2}\eta_{B}}][1-(1-Y_{0})e^{-\frac{\lambda}{2}\eta_{C}}],
\end{aligned}
\end{equation}
where $\lambda\in\{\mu,\nu,\omega,0\}$.
The gain of Bob's conclusive results and Charlie's conclusive results are given by
\begin{equation} \label{yi}
\begin{aligned}
Q_{B\lambda}^{c}=&z[Y_{0}e^{-\frac{\lambda}{2}\eta_{B}}+(\frac{1}{2}+e_{d})(1-e^{-\frac{\lambda}{2}\eta_{B}})]\\
&\times[1-(1-Y_{0})e^{-\frac{\lambda}{2}\eta_{C}}]=P_{B\lambda}^{c}Q_{\lambda},\\
Q_{C\lambda}^{c}=&z[Y_{0}e^{-\frac{\lambda}{2}\eta_{C}}+(\frac{1}{2}+e_{d})(1-e^{-\frac{\lambda}{2}\eta_{C}})]\\
&\times[1-(1-Y_{0})e^{-\frac{\lambda}{2}\eta_{B}}]=P_{C\lambda}^{c}Q_{\lambda}\\
=&\sum_{n=0}^{\infty}e^{-\lambda}\frac{\lambda^{n}}{n!}Y_{Cn},
\end{aligned}
\end{equation}
where $Y_{Cn}$ is the yield given that Alice sends $n$-photon pulse, both Bob's and Charlie's detectors click and Charlie has a conclusive result.
The overall QBER of Bob's conclusive results and Charlie's conclusive results are given by
\begin{equation} \label{er}
\begin{aligned}
e_{B\lambda}^{c}Q_{B\lambda}^{c}=&z[\frac{1}{2}e^{-\frac{\lambda}{2}\eta_{B}}Y_{0}+e_{d}(1-e^{-\frac{\lambda}{2}\eta_{B}})]\\
&\times[1-(1-Y_{0})e^{-\frac{\lambda}{2}\eta_{C}}],\\
e_{C\lambda}^{c}Q_{C\lambda}^{c}=&z[\frac{1}{2}e^{-\frac{\lambda}{2}\eta_{C}}Y_{0}+e_{d}(1-e^{-\frac{\lambda}{2}\eta_{C}})]\\
&\times[1-(1-Y_{0})e^{-\frac{\lambda}{2}\eta_{B}}]\\
=&\sum_{n=0}^{\infty}e^{-\lambda}\frac{\lambda^{n}}{n!}e_{Cn}Y_{Cn},
\end{aligned}
\end{equation}
where $e_{Cn}$ is the QBER of $n$-photon given that Alice sends $n$-photon pulse, both Bob's and Charlie's detectors click and Charlie has a conclusive result. Exploiting the decoy-state method \cite{wang:2005:Beating,Lo:2005:Decoy}, the yield $Y_{C2}$ and QBER $e_{C2}$ of two-photon components in signal-state can be estimated.
The lower bound of $Y_{C2}$ and upper bound of $e_{C2}$ can be given by
\begin{equation} \label{YC2}
\begin{aligned}
Y_{C2}\geq&\frac{2}{\mu\nu\omega(\mu-\nu)(\mu-\omega)(\nu-\omega)}\Big\{\mu\omega(\mu^2-\omega^2)e^{\nu}Q_{C\nu}^{c}\\
&-\mu\nu(\mu^{2}-\nu^{2})e^{\omega}Q_{C\omega}^{c}-\nu\omega(\nu^{2}-\omega^{2})e^{\mu}Q_{C\mu}^{c}\\
&+\big[\mu^{3}(\nu-\omega)+\nu^{3}(\omega-\mu)+\omega^{3}(\mu-\nu)\big]Q_{C0}^{c}\Big\},\\
e_{C2}\leq & \frac{2}{\nu\omega(\nu-\omega)Y_{C2}}\big[\omega e^{\nu}e_{C\nu}^{c}Q_{C\nu}^{c}-\nu e^{\omega}e_{C\omega}^{c}Q_{C\omega}^{c}\\
&-(\omega-\nu) e_{C0}^{c}Q_{C0}^{c}\big].
\end{aligned}
\end{equation}

In the decoy-state method, the finite sample size effect should be taken into account. We exploit the standard error analysis method \cite{Ma:2005:QKD} to calculate the statistical fluctuation.
Thus, we have
\begin{equation} \label{sf}
\begin{aligned}
(Q_{C\lambda}^{c})^{U/L}&=Q_{C\lambda}^{c}\left(1\pm \frac{n_{\alpha1}}{\sqrt{N_{\lambda}Q_{C\lambda}^{c}}}\right),\\
\end{aligned}
\end{equation}
where $N_{\lambda}$ is the number of pulses given that Alice sends weak coherent states with intensity $\lambda$. $N_{\lambda}Q_{C\lambda}^{c}$ is the number of pulses given that Alice sends weak coherent states with intensity $\lambda$ and Charlie has a conclusive result.
Thus, we have $N_{\lambda}=P_{\lambda}N$ and $P_{\lambda}$ is the probability of intensity $\lambda$.

Only the contribution of signal-state can be used as test bits and untested bits. The amount of Charlie's randomly selected test bits is $M_{t}=\beta Q_{\mu}P_{\mu}N$, the remaining bits $M_{u}=(1-\beta)Q_{\mu}P_{\mu}N$ are untested bits.
The distance $\Delta$ is written as
\begin{equation} \label{dis}
\begin{aligned}
\Delta=\Delta_{t}+\delta_{2},~~\Delta_{t}=e_{B\mu}^{c}+e_{C\mu}^{c},\\
\delta_{2}=g[P_{C}^{c}P_{B}^{c}M_{u},P_{C}^{c}P_{B}^{c}M_{t},\Delta_{t},\epsilon_{2}].
\end{aligned}
\end{equation}
Because an unambiguous discrimination among $C$ linearly dependent states of a qubit space is only possible when at least $C-1$ copies of the states are available \cite{Chefles:2001:Unambiguous}.
In the six-state QDS with phase-randomized weak coherent states, Bob cannot unambiguously discriminate the polarization states when Alice sends $3$-photon or $4$-photon pulses. For simplicity, we only consider the contribution of the two-photon component. For vacuum-state and single-photon component, there is a negligible probability to indicate a successful event due to the low dark count rate. Therefore, we can assume that Bob can guess the bits of Charlie's conclusive results without errors unless Alice sends two-photon component pulses.

The optimal probability of Bob accepting the message while Charlie rejecting is
\begin{equation} \label{pro}
\begin{aligned}
\varepsilon_{2}=\textrm{Pr}(BA,CR)=\exp[-\frac{(A-P_{B\mu}^{c}T_{a})^2}{2A}M_{u}],\\
\end{aligned}
\end{equation}
where $A$ is the physical solution of the following equation and inequalities,
\begin{equation} \label{sol}
\begin{aligned}
\frac{(A-P_{B\mu}^{c}T_{a})^2}{2A}&=\frac{\left[P_{C\mu}^{c}T_{v}-P_{C\mu}^{c}\left(A/P_{B\mu}^{c}+\Delta\right)\right]^2}{3P_{C\mu}^{c}\left(A/P_{B\mu}^{c}+\Delta\right)},\\
P_{B\mu}^{c}T_{a}<A&<P_{B\mu}^{c}(T_{v}-\Delta).
\end{aligned}
\end{equation}
The optimal probability of Charlie accepting a forged message is
\begin{equation} \label{pro}
\begin{aligned}
\varepsilon_{1}=\textrm{Pr}(CA)=\exp\left[-\frac{(S_{c}-T_{v2})^2}{2S_{c}}\frac{Q_{C2}^{L}}{Q_{C\mu}^{c}}P_{C\mu}^{c}M_{u}\right],
\end{aligned}
\end{equation}
where $\frac{Q_{C2}^{L}}{Q_{C\mu}^{c}}P_{C\mu}^{c}M_{u}$ is the minimum number of Charlie's conclusive result bits in the untested bits given that Alice sends two-photon component pulses, and
\begin{equation} \label{pro}
\begin{aligned}
T_{v2}=T_{v} \frac{Q_{C\mu}^{c}}{Q_{C2}^{L}},  ~~Q_{C2}^{L}=e^{-\mu}\frac{\mu^{2}}{2}Y_{C2}^{L}.
\end{aligned}
\end{equation}
$T_{v2}$ and $Q_{C2}^{L}$ are the mismatching rate threshold and the gain (lower bound) of the two-photon component, respectively.
The security level of the protocol can be written as
\begin{equation} \label{sol}
\begin{aligned}
\varepsilon_{\textrm{sec}}&=\varepsilon_{\textrm{nor}}+\varepsilon_{\textrm{unf}}\\
&=\varepsilon_{1}+\varepsilon_{2}+\epsilon_{2}+7\epsilon_{3},
\end{aligned}
\end{equation}
where $7\epsilon_{3}$ is the failure probability due to the decoy-state method, and
\begin{equation} \label{fp3}
\begin{aligned}
\epsilon_{3}= \frac{1}{\sqrt{2\pi}}\int_{n_{\alpha1}}^{\infty}e^{-\frac{t^2}{2}}dt,
\end{aligned}
\end{equation}
$n_{\alpha1}$ is the number of standard deviations, we set $n_{\alpha1}=4.753$ for simulation.
The probability of the robustness is
\begin{equation} \label{rob}
\begin{aligned}
\varepsilon_{\textrm{rob}}&=\textrm{Pr}(BR)<\varepsilon'\\
&=h[P_{B\mu}^{c}M_{u},P_{B\mu}^{c}M_{t},e_{B\mu}^{c}, T_{a}-e_{B\mu}^{c}].
\end{aligned}
\end{equation}

\subsection{Practical QDS with Scheme II}

The overall gain $Q_{\gamma\chi}$ can be given by
\begin{equation} \label{gain}
\begin{aligned}
Q_{\gamma\chi}=[1-(1-Y_{0})e^{-\gamma\eta_{B}}][1-(1-Y_{0})e^{-\chi\eta_{C}}],
\end{aligned}
\end{equation}
where $\{\gamma,\chi\}=$ $\{\mu_{1}, \mu_{1}\}$, $\{\mu_{1}, 0\}$, $\{0, \mu_{1}\}$, $\{\nu_{1}, \nu_{1}\}$, $\{\nu_{1}, 0\}$, $\{0, \nu_{1}\}$ and $\{0, 0\}$.
The gain of Bob's conclusive results and Charlie's conclusive results can be written as
\begin{equation} \label{yield}
\begin{aligned}
Q_{B\gamma\chi}^{c}=&z[Y_{0}e^{-\gamma\eta_{B}}+(\frac{1}{2}+e_{d})(1-e^{-\gamma\eta_{B}})]\\
&\times[1-(1-Y_{0})e^{-\chi\eta_{C}}]=P_{B\gamma\chi}^{c}Q_{\gamma\chi},\\
Q_{C\gamma\chi}^{c}=&z[Y_{0}e^{-\chi\eta_{C}}+(\frac{1}{2}+e_{d})(1-e^{-\chi\eta_{C}})]\\
&\times[1-(1-Y_{0})e^{-\gamma\eta_{B}}]=P_{C\gamma\chi}^{c}Q_{\gamma\chi}\\
=&\sum_{n=0}^{\infty}\sum_{m=0}^{\infty}e^{-\gamma}\frac{\gamma^{n}}{n!}e^{-\chi}\frac{\chi^{m}}{m!}Y_{Cnm},
\end{aligned}
\end{equation}
where $Y_{Cnm}$ is the yield that both Bob's and Charlie's detectors click and Charlie has a conclusive result given that Alice sends $n$-photon pulses to Bob and $m$-photon pulses to Charlie.
The overall QBER of Bob's conclusive results and Charlie's conclusive results are given by
\begin{equation} \label{error}
\begin{aligned}
e_{B\gamma\chi}^{c}Q_{B\gamma\chi}^{c}=&z[\frac{1}{2}e^{-\gamma\eta_{B}}Y_{0}+e_{d}(1-e^{-\gamma\eta_{B}})]\\
&\times[1-(1-Y_{0})e^{-\chi\eta_{C}}],\\
e_{C\gamma\chi}^{c}Q_{C\gamma\chi}^{c}=&z[\frac{1}{2}e^{-\chi\eta_{C}}Y_{0}+e_{d}(1-e^{-\chi\eta_{C}})]\\
&\times[1-(1-Y_{0})e^{-\gamma\eta_{B}}]\\
=&\sum_{n=0}^{\infty}\sum_{m=0}^{\infty}e^{-\gamma}\frac{\gamma^{n}}{n!}e^{-\chi}\frac{\chi^{m}}{m!}e_{Cnm}Y_{Cnm},
\end{aligned}
\end{equation}
where $e_{Cnm}$ is the QBER. Exploiting the decoy-state method \cite{wang:2005:Beating,Lo:2005:Decoy}, the yield $Y_{C11}$ and QBER $e_{C11}$ of the two-photon component in the signal-state set can be estimated. It is clear that the estimation of $Y_{C11}$ and $e_{C11}$ is similar to that used in measurement-device-independent QKD \cite{YLT:2014:MDI}. So $Y_{C11}$ and $e_{C11}$ can be written as \cite{YLT:2014:MDI,YF:2015:MQC}
\begin{equation} \label{yield3}
\begin{aligned}
Y_{C11}\geq & \frac{1}{\mu_{1}^{2}\nu_{1}^{2}(\mu_{1}-\nu_{1})}\\
&\times\Big\{\mu_{1}^{3}(e^{2\nu_{1}}Q_{C\nu_{1}\nu_{1}}^{c}-e^{\nu_{1}}Q_{C\nu_{1}0}^{c}-e^{\nu_{1}}Q_{C0\nu_{1}}^{c})\\
&-\nu_{1}^{3}(e^{2\mu_{1}}Q_{C\mu_{1}\mu_{1}}^{c}-e^{\mu_{1}}Q_{C\mu_{1}0}^{c}-e^{\mu_{1}}Q_{C0\mu_{1}}^{c})\\
&+(\mu_{1}^{3}-\nu_{1}^{3})Q_{C00}^{c}\Big\}\\
\end{aligned}
\end{equation}
and
\begin{equation} \label{error3}
\begin{aligned}
e_{C11}\leq & \frac{1}{\nu_{1}^{2}Y_{C11}}\Big(e^{2\nu_{1}}e_{C\nu_{1}\nu_{1}}^{c}Q_{C\nu_{1}\nu_{1}}^{c}-e^{\nu_{1}}e_{C\nu_{1}0}^{c}Q_{C\nu_{1}0}^{c}\\
&-e^{\nu_{1}}e_{C0\nu_{1}}^{c}Q_{C0\nu_{1}}^{c}+e_{C00}^{c}Q_{C00}^{c}\Big).
\end{aligned}
\end{equation}
We exploit the standard error analysis method to calculate the statistical fluctuation.
Thus, we have
\begin{equation} \label{sf}
\begin{aligned}
(Q_{C\gamma\chi}^{c})^{U/L}&=Q_{C\gamma\chi}^{c}\left(1\pm \frac{n_{\alpha2}}{\sqrt{N_{\gamma\chi}Q_{C\gamma\chi}^{c}}}\right).\\
\end{aligned}
\end{equation}
Here, $N_{\gamma\chi}$ is the number of pulses given that Alice sends weak coherent states with intensity set $\{\gamma,\chi\}$.
Thus, $N_{\gamma\chi}=P_{\gamma\chi}N$ and $P_{\gamma\chi}$ is the probability of intensity set $\{\gamma,\chi\}$.

Only the contribution of signal-state set can be used as test bits and untested bits. The amount of Charlie's random selected test bits is $M_{t}=\beta Q_{\mu_{1}\mu_{1}}P_{\mu_{1}\mu_{1}}N$, and the remaining bits $M_{u}=(1-\beta)Q_{\mu_{1}\mu_{1}}P_{\mu_{1}\mu_{1}}N$ are untested bits.
The distance $\Delta$ is written as
\begin{equation} \label{dis}
\begin{aligned}
\Delta&=\Delta_{t}+\delta_{2},~~\Delta_{t}=e_{B\mu_{1}\mu_{1}}^{c}+e_{C\mu_{1}\mu_{1}}^{c},\\
\delta_{2}&=g[P_{C}^{c}P_{B}^{c}M_{u},P_{C}^{c}P_{B}^{c}M_{t},\Delta_{t},\epsilon_{2}].
\end{aligned}
\end{equation}
The optimal probability of Bob accepting the message while Charlie rejecting is
\begin{equation} \label{pro}
\begin{aligned}
\varepsilon_{2}=\textrm{Pr}(BA,CR)=\exp[-\frac{(A-P_{B\mu_{1}\mu_{1}}^{c}T_{a})^2}{2A}M_{u}],\\
\end{aligned}
\end{equation}
where $A$ is the physical solution of the following equation and inequalities,
\begin{equation} \label{sol}
\begin{aligned}
&\frac{(A-P_{B\mu_{1}\mu_{1}}^{c}T_{a})^2}{2A}\\
&=\frac{\left[P_{C\mu_{1}\mu_{1}}^{c}T_{v}-P_{C\mu_{1}\mu_{1}}^{c}\left(A/P_{B\mu_{1}\mu_{1}}^{c}+\Delta\right)\right]^2}{3P_{C\mu_{1}\mu_{1}}^{c}\left(A/P_{B\mu_{1}\mu_{1}}^{c}+\Delta\right)},\\
&P_{B\mu_{1}\mu_{1}}^{c}T_{a}<A<P_{B\mu_{1}\mu_{1}}^{c}(T_{v}-\Delta).
\end{aligned}
\end{equation}
The optimal probability of Charlie accepting a forged message is
\begin{equation} \label{pro}
\begin{aligned}
\varepsilon_{1}&=\textrm{Pr}(CA)\\
&=\exp\left[-\frac{(S_{c}-T_{v11})^2}{2S_{c}}\frac{Q_{C11}^{L}}{Q_{C\mu_{1}\mu_{1}}^{c}}P_{C\mu_{1}\mu_{1}}^{c}M_{u}\right],
\end{aligned}
\end{equation}
where $\frac{Q_{C11}^{L}}{Q_{C\mu_{1}\mu_{1}}^{c}}P_{C\mu_{1}\mu_{1}}^{c}M_{u}$ is the minimum number of Charlie's conclusive result bits in the untested bits given that Alice sends two-photon component pulses, and
\begin{equation} \label{pro}
\begin{aligned}
T_{v11}=T_{v}\frac{Q_{C\mu_{1}\mu_{1}}^{c}}{Q_{C11}^{L}},  ~~Q_{C11}^{L}=e^{-2\mu_{1}}\mu_{1}^{2}Y_{C11}^{L},
\end{aligned}
\end{equation}
$T_{v11}$ and $Q_{C11}^{L}$ are the mismatching rate threshold and the gain (lower bound), respectively.
The security level of the protocol can be written as
\begin{equation} \label{sol}
\begin{aligned}
\varepsilon_{\textrm{sec}}&=\varepsilon_{\textrm{nor}}+\varepsilon_{\textrm{unf}}\\
&=\varepsilon_{1}+\varepsilon_{2}+\epsilon_{2}+11\epsilon_{4}.
\end{aligned}
\end{equation}
where $11\epsilon_{4}$ is the failure probability due to the decoy-state method, and
\begin{equation} \label{fp3}
\begin{aligned}
\epsilon_{4}= \frac{1}{\sqrt{2\pi}}\int_{n_{\alpha2}}^{\infty}e^{-\frac{t^2}{2}}dt,
\end{aligned}
\end{equation}
$n_{\alpha2}$ is the number of standard deviations and we set $n_{\alpha2}=4.845$ for simulation.

The probability of the robustness is
\begin{equation} \label{rob}
\begin{aligned}
\varepsilon_{\textrm{rob}}&=\textrm{Pr}(BR)<\varepsilon'\\
&=h[P_{B\mu_{1}\mu_{1}}^{c}M_{u},P_{B\mu_{1}\mu_{1}}^{c}M_{t},e_{B\mu_{1}\mu_{1}}^{c}, T_{a}-e_{B\mu_{1}\mu_{1}}^{c}].
\end{aligned}
\end{equation}

\subsection{Decoy-state method cannot help the adversary}

For Alice's repudiation attack, the dishonest signer Alice attempts to cheat two honest recipients Bob and Charlie.
Since quantum states are prepared by Alice,  multi-photon component will provide no advantages for Alice.
Only the contribution of signal-state (set) can be use as test bits and untested bits in the QDS with phase-randomized weak coherent states.
The security analysis of repudiation attack does not use the information of the decoy-state.
Therefore, the decoy-state method does not bring any advantage for Alice's repudiation.

In the case of Bob's forgery attack, the dishonest authenticator Bob attempts to cheat the honest signer Alice and the honest verifier Charlie.
The decoy-state method is used to estimate the yield and bit error rate of two-photon component sent by Alice given that Charlie has a conclusive result.
Furthermore, the decoy-state method is used to estimate the minimum number (maximum average information $I_{BC}$) of Charlie's conclusive result bits given that Alice sends two-photon component pulses,
which is used for security against forgery attack.
Thus, Alice and Charlie are honest and they trust each other while Bob is dishonest in the decoy-state method. We recall the decoy-state QKD \cite{wang:2005:Beating,Lo:2005:Decoy} that two communication parties trust each other while the eavesdropper Eve is an untrusted adversary.
Bob is an active participant in the QDS scheme, i.e., he will announce the result whether his detector has a click. Meanwhile, Bob can exploit the insecure quantum channel to decide which qubit (location) has the chance to be detected by Charlie. The above two aspects are equivalent to that Bob can decide the effective event. In the QKD protocol, Eve can also decide which qubit (location) has the chance to be detected by the receiver as the effective event, such as the photon-number-splitting attack \cite{Brassard:2000:PNS}.
Therefore, the decoy-state method cannot bring any advantage for Bob to forge in the QDS.



%

\end{document}